\documentclass[entropy,article,submit,moreauthors,pdftex]{Definitions/mdpi}

\firstpage{1}
\pubvolume{xx}
\issuenum{x}
\articlenumber{x}
\pubyear{2022}
\copyrightyear{2022}
\history{Received: date; Accepted: date; Published: date}

\usepackage{color,xspace}
\usepackage{Definitions/macros_gpi}

        \let\x=\xi        
\let\s=\sigma

\let\P=\Pi         
   
\let\la = \langle \let\ra = \rangle

\def\la{\left\langle}
\def\ra{\right\rangle}
\def\wh{\widehat}

\setitemize{parsep=6pt,itemsep=0pt,leftmargin=*,labelsep=5.5mm}
\setenumerate{parsep=6pt,itemsep=0pt,leftmargin=*,labelsep=5.5mm} 
\setlist[description]{itemsep=0mm}   
\usepackage{environ}
\NewEnviron{myequation}{%
\begin{equation} 
\scalebox{0.95}{$\BODY$}
\end{equation}
}

\Title{Time-Dependent Maximum Entropy Model for Populations of Retinal Ganglion Cells$^\dagger$}

\newcommand{\orcidauthorA}{0000-0001-6217-4370} 
\newcommand{\orcidauthorB}{0000-0002-3131-5537} 

\Author{Geoffroy Delamare$^{1,2}$ orcid number:\orcidauthorA \\ Ulisse Ferrari$^{1}$ orcid number:\orcidauthorB} 

\AuthorNames{Geoffroy Delamare, Ulisse Ferrari}

\address{%
$^1$ \quad Institut de la Vision, Sorbonne Universit\'e, INSERM, CNRS, 17 rue Moreau, 75012, Paris
\\
$^2$ \quad Current address: Bioengineering Department, Imperial College London, London SW7 2AZ, UK
}
\corres{Ulisse Ferrari, ulisse.ferrari@inserm.fr}

\firstnote{Submitted to International Workshop on Bayesian Inference and Maximum Entropy Methods in Science and Engineering, IHP, Paris, July 18-22, 2022.} 

\abstract{
The inverse Ising model is used in computational neuroscience to infer probability distributions of the synchronous activity of large neuronal populations. 
This method allows for finding the Boltzmann distribution with single neuron biases and pairwise interactions that maximizes the entropy and reproduces the empirical statistics of the recorded neuronal activity. 
Here we apply this strategy to large populations of retinal output neurons (ganglion cells) of different types, stimulated by multiple visual stimuli with their own statistics. The activity of retinal output neurons is driven by both the inputs from upstream neurons, which encode the visual information and reflect stimulus statistics, and the recurrent connections, which induce network effects.
We first apply the standard inverse Ising model approach, and show that it accounts well for the system's collective behavior when the input visual stimulus has short-ranged spatial correlations, but fails for long-ranged ones.
This happens because stimuli with long-ranged spatial correlations synchronize the activity of neurons over long distances. This effect cannot be accounted for by pairwise interactions, and so by the pairwise Ising model. 
To solve this issue, we apply a previously proposed framework that includes a temporal dependence in the single neurons biases to model how neurons are driven in time by the stimulus. 
Thanks to this addition, the stimulus effects are taken into account by the biases, and the pairwise interactions allow for characterizing the network effect in the population activity and for reproducing the structure of the recurrent functional connections in the retinal architecture. In particular, the inferred interactions are strong and positive only for nearby neurons of the same type. Inter-type connections are instead small and slightly negative. Therefore, the retinal architecture splits into weakly interacting subpopulations composed of strongly interacting neurons.
Overall, this temporal framework fixes the problems of the standard, static, inverse Ising model and accounts for the system's collective behavior, for stimuli with either short or long-range correlations. 
}

\keyword{\textls[-15]{Inverse problems; Maximum Entropy; Computational Neuroscience; Retinal Ganglion Cells; Neuronal Recordings; Multi-Electrode Array Experiments; Time-Dependent Stimulus Statistics}}

\begin{document}

The inverse Ising model (IM) is a modelling strategy to infer Boltzmann distribution with pairwise interactions from data. 
In systems biology, it has been applied to model the behaviour of large systems with many units that interact one with another, ranging from neuronal ensembles in both early sensory systems  \cite{Schneidman06,Shlens06,Tkacik14,Ferrari18b}, cortex  \cite{Marre09,Hamilton13,Tavoni17,Meshulam17,Donner17,Nghiem18} and neuronal cultures \cite{Schneidman06,Shimazaki15}, to proteins \cite{Weigt09,Santolini14,DeLeonardis15,Figliuzzi16}, antibodies \cite{Mora10} and even flocks of birds \cite{Bialek12}.
To better understand the effectiveness of the inverse IM in modeling biological data, empirical benchmarks \cite{Ferrari17a} and several theoretical investigations \cite{Roudi09,Obuchi15a,Obuchi15b,Merchan16} have also been performed.

The inverse IM approach neglects any temporal evolution of the system and assumes that its activity can be described as a stationary state \cite{Schneidman06}.  
Although this simplification works well in many practical applications, it cannot lead to a satisfying model when the system is strongly driven by external stimuli \cite{Granot-Atedgi13,Nghiem18,Priesemann18,Ferrari18b}. 
In this study, we consider the activity of retinal output neurons in response to visual stimuli with different statistics and show that the inverse IM approach fails in accounting for the empirical statistics when the stimulus has strong and long-ranged correlations.
To solve this issue, the inverse IM framework has been extended to include the effects of time-varying external stimuli into the activity of the retinal output neurons \cite{Granot-Atedgi13}.
More recently \cite{Ferrari18b,Sorochynskyi21}, this time-dependent framework has been empowered by focusing on a population of retinal neurons of the same type.
Here, we perform a step further and consider the case of a population of neurons of two different types, subject to two external stimuli with very different statistics.
Then, in accordance with previous results \cite{Ferrari18b}, we show that the temporal framework provides a very effective model also when the visual stimulus has strong and long-ranged correlations. 

We conclude our work by analysing the properties of the inferred functional interactions between retinal neurons. 
Neurons of the same type are evenly spaced over a two-dimensional triangular lattice, forming regular  \textit{mosaics} \cite{Wassle04}.
The inferred interactions are strong and positive only for nearby neurons of the same type, whereas distant neurons do not interact directly. 
Connections between neurons of different type are instead small (or sometimes slightly negative), also for nearby cells. 
Therefore, the retinal architecture splits into weakly interacting subpopulations of strongly interacting neurons.

\section{Recording of retinal ganglion cells}
\begin{figure}
\centering
\includegraphics[width=\columnwidth]{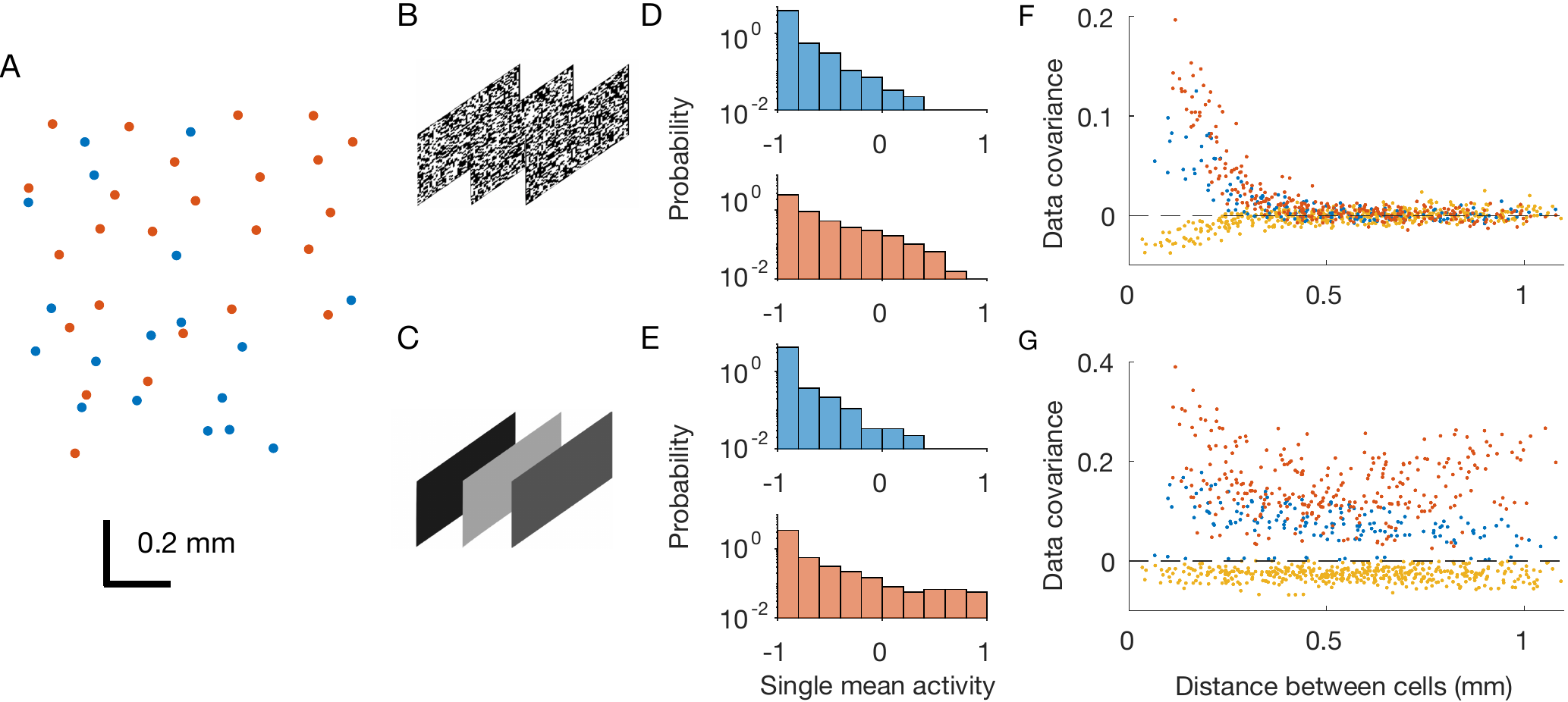}
\caption{\textbf{Retinal multi-electrode array experiments.}
OFF neurons and OFF-OFF pairs are in red, ON and ON-ON pairs in blue,  OFF-ON pairs in yellow.
A) Physical positions of the recorded neurons within the two-dimensional retinal output layer. Each dot correspond to a neuron.
B-C) Two considered stimuli are checkerboard and full-field flicker.
D-E) Distribution of single neuron mean activities.
F-G) Pairwise covariances as a function of the physical distance between the neurons.
}
\label{f:1}
\end{figure}

We focus on the activity of two populations of 18 ON and 25 OFF rat retinal output neurons (known as ganglion cells) \cite{Deny17}, recorded during one \textit{ex-vivo} multi-electrode array experiment \cite{Marre12}.
These experiments allow to measure the times at which each neuron emits a spike in response to an ongoing visual stimulation. 
ON and OFF neurons have opposite polarities, meaning that they respond preferentially to, respectively, increase or decrease of light intensity \cite{Wassle04}.
Additionally, thanks to standard techniques \cite{Marre12}, it is possible to locate the position of each neuron within the two-dimensional retinal output layer (Fig. \ref{f:1}A).
To validate our results, we also consider a second experiment where 21 ON and 32 OFF retinal output neurons were stimulated with the same videos.

During the experiments, the retina was stimulated by two different black-and-white videos repeated multiple times (Fig. \ref{f:1}B\&C):  
a white-noise checkerboard stimulus with strong but short-ranged spatial correlations, and a full-field video whose luminosity flickers over different grey values, \textit{i.e.} with strong spatial correlations that extend over the entire scene. 

After binning the spiking activity with small windows of $\Delta t = 20ms$, we can associate to each neuron $i$ in each time-bin $t$ during repetition $r$ a binary variable $\sigma_i^{r}(t)$ equal to $+1$ if the neuron spiked in the time-bin or $-1$ if not. 
Thanks to this preprocessing, we end up with a sequence of snapshots of neuronal activity $\{  \s_i^r(t) \}_{i=1}^N$, which can be seen as observations of system configurations. 
At first, we estimated each neuron's \textit{mean activity}, that is the average of $\sigma_i^{r}(t)$ over the recordings. 
Mean activities in response to the two stimuli were very similar (Fig. \ref{f:1}D\&E), for both type of neurons. 
However, covariances were different across stimuli: the checkerboard induced strong, short-ranged correlations whereas the full-field induced strong correlations over longer distances. (Fig. \ref{f:1}F\&G).

\section{Inverse disordered Ising model}
In order to analyse the retinal spiking activity, Schneidman et al. \cite{Schneidman06} has proposed to consider the probability distribution $P(\bm \s)$ of observing a given activity snapshot  $\bm \s$, regardless of the time at which it has been observed.
As shown before (Fig.\ref{f:1}F and G), neuronal activities show strong correlations, suggesting that neurons are not independent. 
Therefore, $P(\bm \s)$ can not be modelled as a collection of independent distributions, but it requires an interacting model. 
For this scope, the principle of maximum entropy suggests to consider all the probability distributions reproducing the empirical mean of all the single variable terms ($\s_i$) and their pairwise products ($\s_i\s_j$), the covariances, to then select the one with the largest entropy. 
This leads to the construction of the well known pairwise disordered Ising model (IM) \cite{Schneidman06,Cocco11}:

\begin{equation}
P( {\bm \s}) \sim \exp \Big\{  \sum_i h_i \s_i + \sum_{i<j}J_{ij} \s_i \s_j   \Big \} ~,
\label{intro:boltzDistr}
\end{equation}

with yet unknown biases ${\bm h}$ and couplings ${\bm J}$, that have to be inferred from data. 
To estimate these parameters, we can compute the model (log-)likelihood over the dataset, and search for the set of parameters that maximises it \cite{Cocco11}.
Additionally, in order to limit noise effects, we added an $L_2$ regularization over the biases ${\bm h}$ and an $L_1$ regularization over the couplings ${\bm J}$ \cite{Cocco11}.
Finally, because the considered systems are too large for performing an exact inference, we used a pseudo-Newton Markov-chain Monte-Carlo algorithm \cite{Ferrari16}.

\begin{figure}
\centering
\includegraphics[width=\columnwidth]{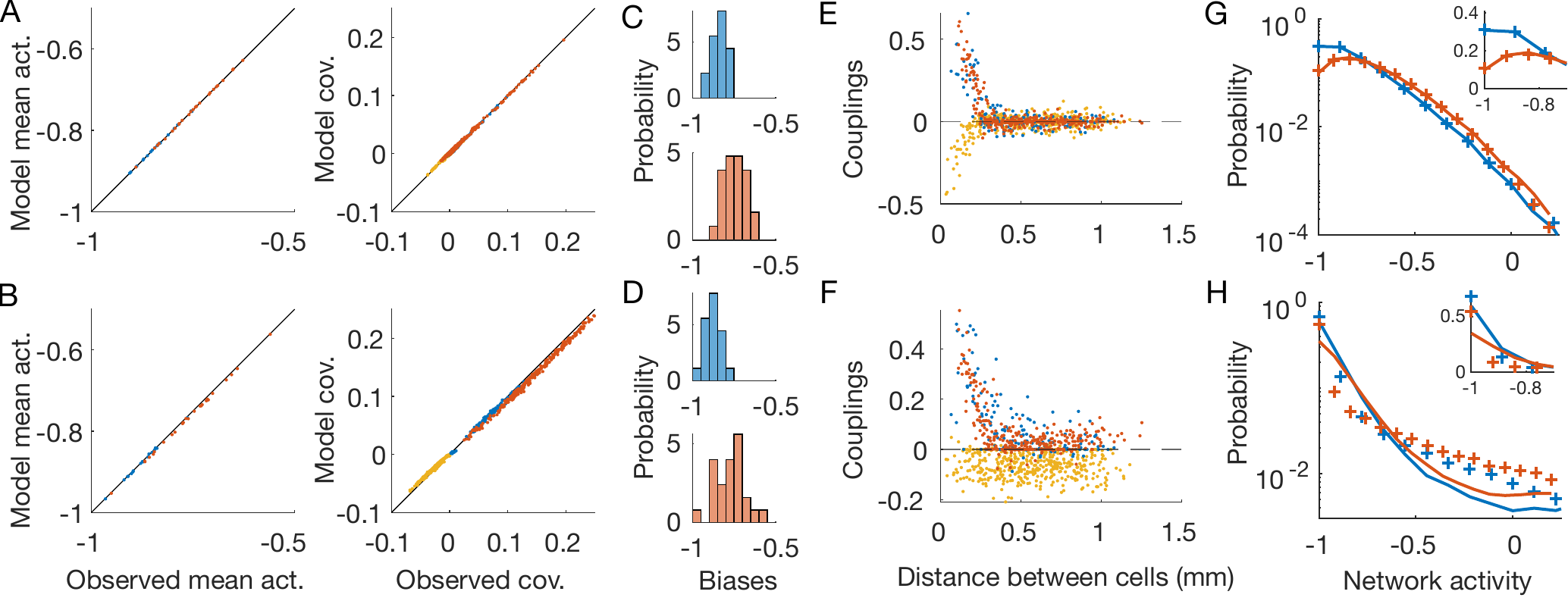}
\caption{\textbf{Inverse disordered Ising model.}
OFF neurons and OFF-OFF pairs are in red, ON and ON-ON in blue,  OFF-ON pairs in yellow. 
The first line shows results for the checkerboard stimulus, the second for the full-field one (Fig.\ref{f:1}B and C).
A,B) Inferred IM reproduces the neurons' mean activities and covariances with high precision.
C,D) Distribution of the inferred biases ${\bm h}$.
E,F) Inferred pairwise couplings ${\bm J}$ as a function of the physical distance between the neurons.
G,H) Empirical (plus signs) and model-predicted (lines) probability distributions of the network activity of the two neuronal populations. Inset: zoom in linear scale.
}
\label{f:2}
\end{figure} 

As expected by model construction \cite{Cocco11}, the inferred distributions were able to reproduce the neurons' mean activities and covariances (Fig. \ref{f:2}A, B), showing that we solved the inference problem for both stimuli with high accuracy.
Both biases (Fig. \ref{f:2}C, D) and couplings (Fig. \ref{f:2}E, F) inferred from the response to the two stimuli are different. 
In particular, for the checkerboard stimulus, which has short-ranged stimulus correlations, we observe strong positive or negative couplings only between nearby neurons, while couplings between distant ones are very small. 
For the full-field video, which instead has long-ranged stimulus correlations, we observe strong couplings, even at large distances.
Overall these results show that the inferred couplings depend on the correlation structure of the stimulus: by acting as correlated input to the neurons, the stimulus induces strong correlations among certain pairs of neurons, and consequently strong couplings among them \cite{Ferrari18b}.

Lastly, we notice how the inferred inverse disordered IM is capable of predicting the empirical probability distribution of the network activity ($\sum_i \sigma_i$) for the checkerboard stimulus, but it fails to do so for the full-field flicker (Fig. \ref{f:2}F,G).
This distribution reflects the collective behaviour of the whole system, and therefore depends on the high-order statistics of the neuronal activities. 
As such, the pairwise structure of the checkerboard video, due to the short-ranged correlations, can be accounted for by a model with pairwise couplings.
However, for the full-field flicker, the stimulus synchronises the whole neuronal population altogether. As a consequence, the correlations structure is not pairwise and the pairwise inverse IM struggles to reproduce such higher-order neuronal statistics. 
A similar effect has been reported previously for the activity of cortical neurons during Slow-Wave Sleep \cite{Nghiem18}.

\section{Time-dependent model}
Instead of constructing a single probability distribution $P({\bm \s})$ for the whole recording, in the time-dependent Ising model framework \cite{Granot-Atedgi13, Ferrari18b}, we build a collection of probability distributions $ \{ P^t({\bm \s} ) \}_{t=1}^T$, one for each time-bin.
Following the maximum entropy principle, we search for the probability distribution that has the maximum entropy among those that reproduce the mean single neuron activities in each time-bin $\la \s_i(t) \ra = 1/R \sum_r \s_i^r(t)$, where $r$ runs over the $R$ repetitions of the stimulus. 
We also require that the model reproduces the total pairwise correlations $\la \s_i \s_j \ra = 1/(RT) \sum_{r,t} \s_i^r(t) \s_j^r(t)$ computed over both time and repetitions, the same observables imposed for the inverse IM (Eq. \ref{intro:boltzDistr}).
This leads us to the following model:

\begin{equation}
P_t({\bm \s} ) \sim \exp \Big\{ \sum_i h_i[t] \s_i  + \sum_{i<j} J_{ij} \s_i  \s_j \Big\}~.
\label{eqIsingTemporal}
\end{equation}

In the model (\ref{eqIsingTemporal}) the biases ${\bm h[t]}$ carry the temporal dependence that accounts for the time-evolution of the stimulus drive. 
However, because we haven't asked the model to reproduce the pairwise correlations in each temporal window, but only the averaged one, the couplings ${\bm J}$ are constant in time.
This choice is biologically motivated: the couplings reflect the internal connections between neurons within the retinal architecture and therefore should be independent of the stimulus \cite{Ferrari18b}.
Additionally, this also limits the number of parameters avoiding the risk of overfitting. 
As in the inverse IM, we include an $L_2$ regularisation on the biases and an $L_1$ on the couplings with the same strength.

\begin{figure}
\centering
\includegraphics[width=\columnwidth]{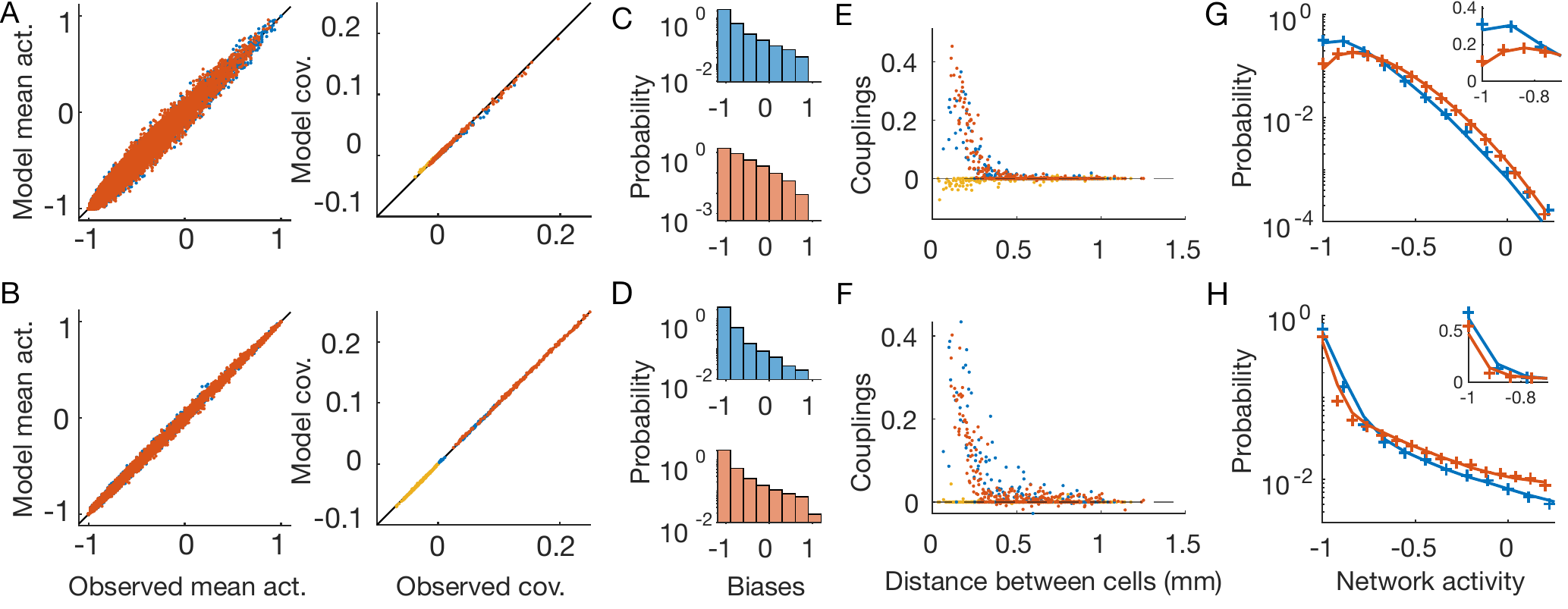}
\caption{\textbf{Time-dependent Ising model.}
OFF neurons and OFF-OFF pairs are in red, ON and ON-ON in blue,  OFF-ON pairs in yellow. 
The first line show results for the checkerboard stimulus, the second for the full-field one (Fig.\ref{f:1}B and C).
A,B) Inferred time-dependent IM reproduces the neurons' mean activities and covariances with high precision.
C,D) Distribution of the inferred biases ${\bm h[t]}$.
E,F) Inferred pairwise couplings ${\bm J}$ as a function of the physical distance between the neurons.
G,H) Empirical (plus signs) and model-predicted (lines) probability distributions of the network activity of the two neuron populations. Inset: zoom in linear scale.
}
\label{f:3}
\end{figure} 

As expected by model construction, the inferred time-dependent distributions reproduce the empirical mean activities and covariances (Fig. \ref{f:3}A, B), showing that we solved the inference problem for both stimuli with high accuracy.
As before (Fig. \ref{f:2}C,D), the inferred biases show different distributions for the two stimuli (Fig. \ref{f:2}C,D). 
The inferred couplings have instead a much similar behavior (Fig. \ref{f:3}E,F), showing a fast decay with the distance between the neuron pairs, for both the checkerboard and the full-field stimulus. 
In particular, those between neurons of different types are zero or slightly negative, whereas those between nearby neurons of the same type large and positive.
Lastly, in the case of the time-dependent IM, the inferred model is capable of predicting  the empirical probability distribution of the network activity for both stimuli with high accuracy (Fig. \ref{f:3}G,H).
Consistently with previous findings \cite{Ferrari18b}, these results show that by using time-dependent IM we are capable of disentangling the collective behaviours that arise because neurons receive correlated inputs, from those that are instead due to network effects.

\section{The geometry of the functional connectivity}
\begin{figure}
\centering
\includegraphics[width=\columnwidth]{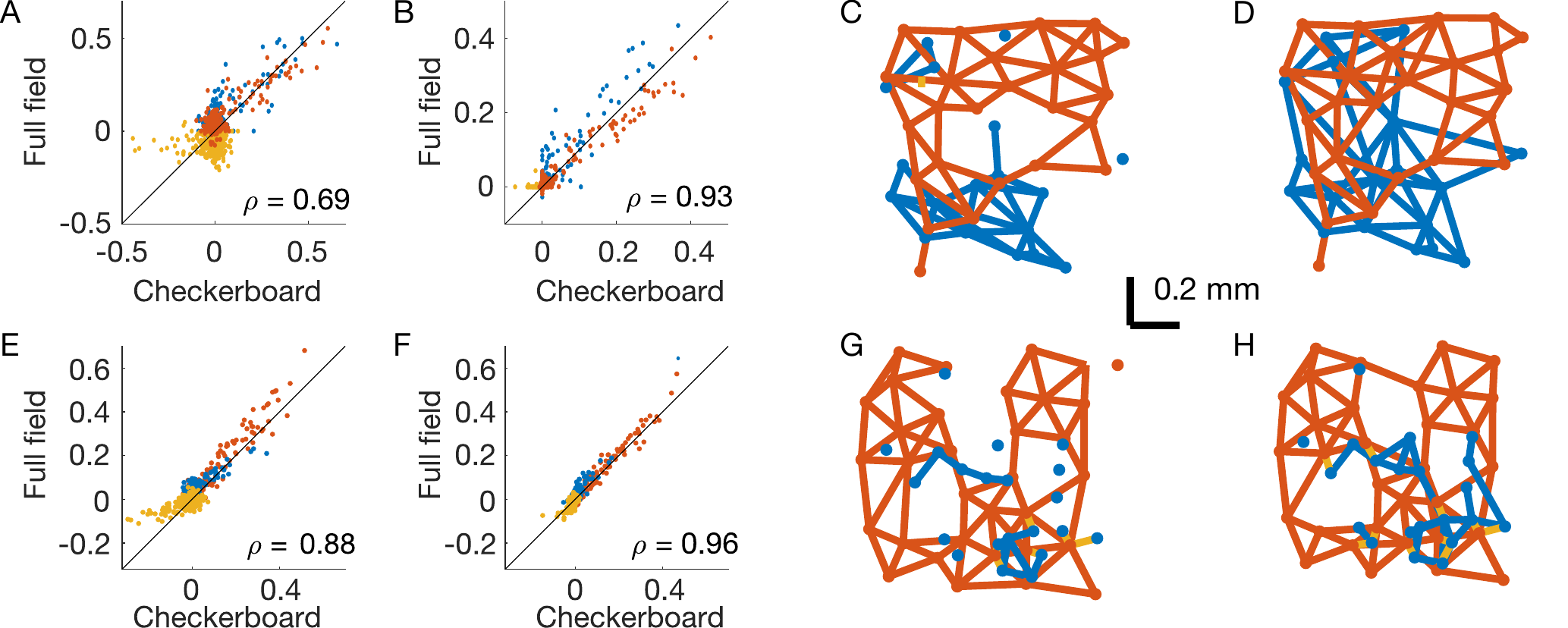}
\caption{\textbf{Structure of the inferred couplings.}
OFF-OFF couplings are in red, ON-ON in blue,  OFF-ON in yellow.
A) Scatterplot of the inferred couplings for the two stimuli in the inverse IM.
B) As (A), but for the time-dependent IM.
$\rho$ is the Pearson correlation.
C,D) Interaction lattice after thresholding the inferred couplings of the time-dependent IM for the checkerboard (C) and full-field stimuli (D).
E-H, same as A-D, but for a second example experiment where retinal neurons responded to the same visual stimulations.
}
\label{f:4}
\end{figure} 

The behaviour of the inferred couplings with distance from the response to the full-field stimulus (Figs. \ref{f:2}F and \ref{f:3}F) are very different.
In the case of the inverse IM, couplings are strong also for distant pairs and seem to reflect the correlation structure of the stimulus. 
In the case of the time-dependent IM, instead, interactions decrease fast with distance and seem not to reflect the correlation structure of the stimulus.
To test for this, we compare the couplings inferred from the two stimuli (Fig.\ref{f:4} A,B). 
In the case of the time-dependent IM, the couplings are indeed much more similar (Pearson correlation $\rho = 0.935$, against $\rho = 0.699$ for the inverse IM).
We conclude that the inferred couplings of the time-dependent IM reflect only the functional connectivity between retinal output neurons.

Retinal output neurons lie on a two-dimensional layer, and their positions can easily be determined by standard methods as their receptive field centres \cite{Marre12} (Fig.\ref{f:1}A).
In order to better visualise the structure of the inferred couplings of the time-dependent IM, we can introduce an arbitrary, but robust, small threshold, set to zero all the smaller couplings ($|J|<0.05$), and draw an interaction lattice (Fig. \ref{f:4}C, D).
After thresholding, the lattice splits into two subcomponents, one for each type, with mostly nearest-neighbour interactions.
Unfortunately, during these experiments, it is difficult to detect all the neurons of a given type within the recorded retinal patch. 
Some neurons are therefore missing, and this prevents a solid study of the lattice connectivity. 
However, given well-known results on retinal mosaics of the literature \cite{Wassle04}, and by looking at the most complete region of (Fig. \ref{f:4}C, D), we expect that if we were able to record all the neurons, the resulting lattice would be an irregular honeycomb, with connectivity equal to six.
Inferred interactions are strong and positive only for nearby neurons of the same type (Fig.\ref{f:3}E,F).
Consequently, only couplings between nearby neurons are above threshold and the functional connectivity lattice show nearest neighbour interactions.
Additionally, because inter-type connections are very small or slightly negative, the retinal architecture splits into weakly interacting subpopulations - each composed of strongly interacting neurons. 

In order to corroborate these results, we have performed the same analysis on a second example dataset where retinal neurons were stimulated with the same visual stimulations (both checkerboard and full-field).
Results are fully consistent and very similar to those of the first experiment (Fig.\ref{f:4}E-H).

\section{Conclusions}
In this work, we focused on modelling the activity of two large populations of retinal output neurons of different types. 
We inferred two different models: the widely used disordered IM \cite{Schneidman06}, and its more recent development, the time-dependent IM \cite{Granot-Atedgi13,Ferrari18b,Sorochynskyi21}.
For each model, we compared the inferred parameters obtained from the retinal response to two very different visual stimulations: the checkerboard, with short spatial correlations, and the full-field with long ones.
In particular we showed that the inferred couplings of the disordered IM, but not those of the time-dependent IM, depend strongly on the stimulus statistics (Fig.\ref{f:4}A,E against Fig.\ref{f:4}B,F).
Consistently, the inferred couplings of the second model are very similar across visual stimulations (Fig.\ref{f:4}B, F).
The time-dependent model is therefore capable of disentangling the collective behaviours induced by the correlated inputs to the retinal output neurons, from those arising from network effects \cite{Ferrari18b}.
As a consequence, we can interpret the inferred couplings as functional connections and characterise the structure of the retinal output-layer architecture.
The inferred interactions are strong and positive only for nearby neurons of the same type, whereas distant neurons do not interact directly. 
Connections between neurons of different type are instead small (or sometimes slightly negative), also for nearby cells. 
Therefore, the retinal architecture splits into weakly interacting subpopulations composed of strongly interacting neurons.

In principle, functional connectivity could also be obtained from the disordered IM inferred from spontaneous activity where a constant, full-field stimulus is played.
In this case, the stimulus has no spatial correlations and the inferred couplings will only reflect the structure of the retinal connectivity.
However, in wildtype retinas, spontaneous activity is usually very weak (few $Hz$) compared to stimulated activity (up to 50-60 $Hz$ in our case).
As a consequence, in order to have the same empirical statistics, one would need much longer recording, which are however very difficult to obtain because of experimental instabilities and limitations.

The inferred functional connectivity matches with known properties of biological networks. 
Depending on the type, output retinal neurons can be connected by direct gap-junction between nearby pairs, or by an indirect connection through multiple gap-junction passing through amacrine cells \cite{Brivanlou98}.
In both cases, only nearby neurons are strongly interacting, and this is nicely reproduced by the functional connections inferred from the time-dependent model.
Network effects can also arise from shared noise coming from presynaptic neurons in the retina - mostly photoreceptors noise.
This effect can explain the negative correlations between neurons of different type (and polarity) \cite{Volgyi09}, and in turn the slightly negative couplings inferred for nearby neurons.

The time-dependent IM takes into account the stimulus effects with the temporal dependence of the single neuron biases ${\bm h[t]}$. 
Instead of modelling the stimulus processing performed by the retina directly, it only reproduces the response behavior in time.
Consistently, in order to infer the model, we only used the response to repeated stimulations, without the need of the actual videos.
As a consequence, the time-dependent IM cannot generalize to new \textit{unseen} (during training) stimuli and this might limit its possible applications.
To overcome these limitations, the time-dependent IM has been extended to its stimulus-dependent generalization \cite{Granot-Atedgi13,Ferrari18b}, where the biases become actual functions of the stimulus.
This allows for inferring deep convolutional neural networks \cite{Deny17,McIntosh16,Mahuas20,Goldin21} to predict the mean neuronal response to stimulus, combined with IM couplings to account for network effects.

\section{Acknowledgments}
This work was supported by the Agence Nationale de la Recherche (ANR-21-CE37-0024 NatNetNoise), by LabEx LIFESENSES (ANR-10-LABX-65), by IHU FOReSIGHT (ANR-18-IAHU-01, IHU-AIDE-UF), by Sorbonne Université with the Emergence program (CrInforNet) and by a grant from AVIESAN-UNADEV (AIDE).

\reftitle{References}


\begin{thebibliography}{-------}
\providecommand{\natexlab}[1]{#1}

\bibitem[Schneidman \em{et~al.}(2006)Schneidman, Berry, Segev, and
  Bialek]{Schneidman06}
Schneidman, E.; Berry, M.; Segev, R.; Bialek, W.
\newblock \em{Weak pairwise correlations imply strongly correlated network
  states in a population }\em.
\newblock {\em \em Nature \em} {\bf 2006}, {\em {\bf 440}},~1007.

\bibitem[Shlens \em{et~al.}(2006)Shlens, Field, Gauthier, Grivich, Petrusca,
  Sher, Litke, and Chichilnisky]{Shlens06}
Shlens, J.; Field, G.D.; Gauthier, J.L.; Grivich, M.I.; Petrusca, D.; Sher, A.;
  Litke, A.M.; Chichilnisky, E.
\newblock The structure of multi-neuron firing patterns in primate retina.
\newblock {\em Journal of Neuroscience} {\bf 2006}, {\em 26},~8254--8266.

\bibitem[Tkacik \em{et~al.}(2014)Tkacik, Marre, Amodei, Schneidman, Bialek, and
  M.J]{Tkacik14}
Tkacik, G.; Marre, O.; Amodei, D.; Schneidman, E.; Bialek, W.; M.J, B.
\newblock \em{Searching for collective behaviour in a network of real neurons
  }\em.
\newblock {\em \em PloS Comput. Biol.\em} {\bf 2014}, {\em {\bf
  10(1)}},~e1003408.

\bibitem[Ferrari \em{et~al.}(2018)Ferrari, Deny, Chalk, Tka{\v{c}}ik, Marre,
  and Mora]{Ferrari18b}
Ferrari, U.; Deny, S.; Chalk, M.; Tka{\v{c}}ik, G.; Marre, O.; Mora, T.
\newblock Separating intrinsic interactions from extrinsic correlations in a
  network of sensory neurons.
\newblock {\em Physical Review E} {\bf 2018}, {\em 98},~042410.

\bibitem[Marre \em{et~al.}(2009)Marre, El~Boustani, Fr\'egnac, and
  Destexhe]{Marre09}
Marre, O.; El~Boustani, S.; Fr\'egnac, Y.; Destexhe, A.
\newblock Prediction of Spatiotemporal Patterns of Neural Activity from
  Pairwise Correlations.
\newblock {\em Phys. Rev. Lett.} {\bf 2009}, {\em 102},~138101.

\bibitem[Hamilton \em{et~al.}(2013)Hamilton, Sohl-Dickstein, Huth, Carels,
  Deisseroth, and Bao]{Hamilton13}
Hamilton, L.S.; Sohl-Dickstein, J.; Huth, A.G.; Carels, V.M.; Deisseroth, K.;
  Bao, S.
\newblock \em{Optogenetic Activation of an Inhibitory Network Enhances
  Feedforward Functional Connectivity in Auditory Cortex}\em.
\newblock {\em \em Neuron \em} {\bf 2013}, {\em {\bf 80}},~1066--76.

\bibitem[Tavoni \em{et~al.}(2017)Tavoni, Ferrari, Battaglia, Cocco, and
  Monasson]{Tavoni17}
Tavoni, G.; Ferrari, U.; Battaglia, F.; Cocco, S.; Monasson, R.
\newblock Functional Coupling Networks Inferred from Prefrontal Cortex Activity
  Show Experience-Related Effective Plasticity.
\newblock {\em Network Neuroscience} {\bf 2017}, pp. 1--27.

\bibitem[Meshulam \em{et~al.}(2017)Meshulam, Gauthier, Brody, Tank, and
  Bialek]{Meshulam17}
Meshulam, L.; Gauthier, J.L.; Brody, C.D.; Tank, D.W.; Bialek, W.
\newblock Collective behavior of place and non-place neurons in the hippocampal
  network.
\newblock {\em Neuron} {\bf 2017}, {\em 96},~1178--1191.

\bibitem[Donner \em{et~al.}(2017)Donner, Obermayer, and Shimazaki]{Donner17}
Donner, C.; Obermayer, K.; Shimazaki, H.
\newblock Approximate inference for time-varying interactions and macroscopic
  dynamics of neural populations.
\newblock {\em PLoS computational biology} {\bf 2017}, {\em 13},~e1005309.

\bibitem[Nghiem \em{et~al.}(2018)Nghiem, Telenczuk, Marre, Destexhe, and
  Ferrari]{Nghiem18}
Nghiem, T.A.; Telenczuk, B.; Marre, O.; Destexhe, A.; Ferrari, U.
\newblock Maximum-entropy models reveal the excitatory and inhibitory
  correlation structures in cortical neuronal activity.
\newblock {\em Physical Review E} {\bf 2018}, {\em 98},~012402.

\bibitem[Shimazaki \em{et~al.}(2015)Shimazaki, Sadeghi, Ishikawa, Ikegaya, and
  Toyoizumi]{Shimazaki15}
Shimazaki, H.; Sadeghi, K.; Ishikawa, T.; Ikegaya, Y.; Toyoizumi, T.
\newblock Simultaneous silence organizes structured higher-order interactions
  in neural populations.
\newblock {\em Scientific reports} {\bf 2015}, {\em 5},~9821.

\bibitem[Weigt \em{et~al.}(2009)Weigt, White, Szurmant, Hoch, and Hwa]{Weigt09}
Weigt, M.; White, R.; Szurmant, H.; Hoch, J.; Hwa, T.
\newblock \em{ Identification of direct residue contacts in protein–protein
  interaction by message passing }\em.
\newblock {\em \em PNAS\em} {\bf 2009}, {\em {\bf 106(1)}},~67--72.

\bibitem[Santolini \em{et~al.}(2014)Santolini, Mora, and Hakim]{Santolini14}
Santolini, M.; Mora, T.; Hakim, V.
\newblock \em{A General Pairwise Interaction Model Provides an Accurate
  Description of In Vivo Transcription Factor Binding Sites. }\em.
\newblock {\em \em PLoS Comput Biol \em} {\bf 2014}, {\em {\bf 9(6)}},~E99015.

\bibitem[{De Leonardis} \em{et~al.}(2015){De Leonardis}, Lutz, Ratz, Cocco,
  Monasson, Schug, and Weigt]{DeLeonardis15}
{De Leonardis}, E.; Lutz, B.; Ratz, S.; Cocco, S.; Monasson, R.; Schug, A.;
  Weigt, M.
\newblock {Direct-Coupling Analysis of nucleotide coevolution facilitates RNA
  secondary and tertiary structure prediction}.
\newblock {\em Nucleic Acids Res.} {\bf 2015}, {\em 43},~10444--10455.
\newblock
  doi:{\changeurlcolor{black}\href{https://doi.org/10.1093/nar/gkv932}{\detokenize{10.1093/nar/gkv932}}}.

\bibitem[Figliuzzi \em{et~al.}(2016)Figliuzzi, Jacquier, Schug, Tenaillon, and
  Weigt]{Figliuzzi16}
Figliuzzi, M.; Jacquier, H.; Schug, A.; Tenaillon, O.; Weigt, M.
\newblock {Coevolutionary landscape inference and the context-dependence of
  mutations in beta-lactamase tem-1}.
\newblock {\em Mol. Biol. Evol.} {\bf 2016}, {\em 33},~268--280,
  \href{http://xxx.lanl.gov/abs/1510.03224}{{\normalfont [1510.03224]}}.
\newblock
  doi:{\changeurlcolor{black}\href{https://doi.org/10.1093/molbev/msv211}{\detokenize{10.1093/molbev/msv211}}}.

\bibitem[Mora \em{et~al.}(2010)Mora, Walczak, Bialek, and Callan]{Mora10}
Mora, T.; Walczak, A.M.; Bialek, W.; Callan, C.G.
\newblock {Maximum entropy models for antibody diversity.}
\newblock {\em Proc. Natl. Acad. Sci.} {\bf 2010}, {\em 107},~5405--5410,
  \href{http://xxx.lanl.gov/abs/0912.5175}{{\normalfont [0912.5175]}}.
\newblock
  doi:{\changeurlcolor{black}\href{https://doi.org/10.1073/pnas.1001705107}{\detokenize{10.1073/pnas.1001705107}}}.

\bibitem[Bialek \em{et~al.}(2012)Bialek, Cavagna, Giardina, Mora, Silvestri,
  Viale, and Walczak]{Bialek12}
Bialek, W.; Cavagna, A.; Giardina, I.; Mora, T.; Silvestri, E.; Viale, M.;
  Walczak, A.M.
\newblock {Statistical mechanics for natural flocks of birds.}
\newblock {\em Proc. Natl. Acad. Sci. U. S. A.} {\bf 2012}, {\em
  109},~4786--91,  \href{http://xxx.lanl.gov/abs/1107.0604}{{\normalfont
  [1107.0604]}}.
\newblock
  doi:{\changeurlcolor{black}\href{https://doi.org/10.1073/pnas.1118633109}{\detokenize{10.1073/pnas.1118633109}}}.

\bibitem[Ferrari \em{et~al.}(2017)Ferrari, Obuchi, and Mora]{Ferrari17a}
Ferrari, U.; Obuchi, T.; Mora, T.
\newblock Random versus maximum entropy models of neural population activity.
\newblock {\em Phys. Rev. E} {\bf 2017}, {\em 95},~042321.

\bibitem[Roudi \em{et~al.}(2009)Roudi, Nirenberg, and Latham]{Roudi09}
Roudi, Y.; Nirenberg, S.; Latham, P.E.
\newblock Pairwise maximum entropy models for studying large biological
  systems: when they can work and when they can't.
\newblock {\em PLoS computational biology} {\bf 2009}, {\em 5},~e1000380.

\bibitem[Obuchi \em{et~al.}(2015)Obuchi, Cocco, and Monasson]{Obuchi15a}
Obuchi, T.; Cocco, S.; Monasson, R.
\newblock Learning probabilities from random observables in high dimensions:
  the maximum entropy distribution and others.
\newblock {\em Journal of Statistical Physics} {\bf 2015}, {\em 161},~598--632.

\bibitem[Obuchi and Monasson(2015)]{Obuchi15b}
Obuchi, T.; Monasson, R.
\newblock Learning probability distributions from smooth observables and the
  maximum entropy principle: some remarks.
\newblock  Journal of Physics: Conference Series. IOP Publishing, IOP
  Publishing,  2015, Vol. 638, p. 012018.

\bibitem[Merchan and Nemenman(2016)]{Merchan16}
Merchan, L.; Nemenman, I.
\newblock On the Sufficiency of Pairwise Interactions in Maximum Entropy Models
  of Networks.
\newblock {\em Journal of Statistical Physics} {\bf 2016}, {\em
  162},~1294--1308.

\bibitem[Granot-Atedgi \em{et~al.}(2013)Granot-Atedgi, Tkacik, Segev, and
  Schneidman]{Granot-Atedgi13}
Granot-Atedgi, E.; Tkacik, G.; Segev, R.; Schneidman, E.
\newblock Stimulus-dependent Maximum Entropy Models of Neural Population Codes.
\newblock {\em PLOS Computational Biology} {\bf 2013}, {\em 9},~1--14.
\newblock
  doi:{\changeurlcolor{black}\href{https://doi.org/10.1371/journal.pcbi.1002922}{\detokenize{10.1371/journal.pcbi.1002922}}}.

\bibitem[Priesemann and Shriki(2018)]{Priesemann18}
Priesemann, V.; Shriki, O.
\newblock Can a time varying external drive give rise to apparent criticality
  in neural systems?
\newblock {\em PLoS computational biology} {\bf 2018}, {\em 14},~e1006081.

\bibitem[Sorochynskyi \em{et~al.}(2021)Sorochynskyi, Deny, Marre, and
  Ferrari]{Sorochynskyi21}
Sorochynskyi, O.; Deny, S.; Marre, O.; Ferrari, U.
\newblock Predicting synchronous firing of large neural populations from
  sequential recordings.
\newblock {\em PLoS computational biology} {\bf 2021}, {\em 17},~e1008501.

\bibitem[W\"assle(2004)]{Wassle04}
W\"assle, H.
\newblock \em{Parallel processing in the mammalian retina}\em.
\newblock {\em \em Nature Reviews Neuroscience\em} {\bf 2004}, {\em {\bf
  5}},~747--57.

\bibitem[Deny \em{et~al.}(2017)Deny, Ferrari, Mace, Yger, Caplette, Picaud,
  Tka{\v{c}}ik, and Marre]{Deny17}
Deny, S.; Ferrari, U.; Mace, E.; Yger, P.; Caplette, R.; Picaud, S.;
  Tka{\v{c}}ik, G.; Marre, O.
\newblock Multiplexed computations in retinal ganglion cells of a single type.
\newblock {\em Nature communications} {\bf 2017}, {\em 8},~1964.

\bibitem[Marre \em{et~al.}(2012)Marre, Amodei, Deshmukh, Sadeghi, Soo, Holy,
  and Berry]{Marre12}
Marre, O.; Amodei, D.; Deshmukh, N.; Sadeghi, K.; Soo, F.; Holy, T.; Berry, M.
\newblock \em{Recording of a large and complete population in the retina}\em.
\newblock {\em \em Journal of Neuroscience \em} {\bf 2012}, {\em {\bf
  32(43)}},~1485973.

\bibitem[Cocco and Monasson(2011)]{Cocco11}
Cocco, S.; Monasson, R.
\newblock \em{Adaptive cluster expansion for inferring Boltzmann machines with
  noisy data}\em.
\newblock {\em \em Phys. Rev. Lett. \em} {\bf 2011}, {\em {\bf 106}},~090601.

\bibitem[Ferrari(2016)]{Ferrari16}
Ferrari, U.
\newblock Learning maximum entropy models from finite-size data sets: A fast
  data-driven algorithm allows sampling from the posterior distribution.
\newblock {\em Phys. Rev. E} {\bf 2016}, {\em 94},~023301.

\bibitem[Brivanlou \em{et~al.}(1998)Brivanlou, Warland, and
  Meister]{Brivanlou98}
Brivanlou, I.; Warland, D.; Meister, M.
\newblock \em{Mechanisms of Concerted Firing among Retinal Ganglion Cells}\em.
\newblock {\em \em Neuron \em} {\bf 1998}, {\em {\bf 20}},~527--539.

\bibitem[V\"{o}lgyi \em{et~al.}(2009)V\"{o}lgyi, Chheda, and
  Bloomfield]{Volgyi09}
V\"{o}lgyi, B.; Chheda, S.; Bloomfield, S.
\newblock Tracer Coupling Patterns of the Ganglion Cell Subtypes in the Mouse
  Retina.
\newblock {\em J. Comp. Neurol.} {\bf 2009}, {\em 512},~664--687.

\bibitem[McIntosh \em{et~al.}(2016)McIntosh, Maheswaranathan, Nayebi, Ganguli,
  and Baccus]{McIntosh16}
McIntosh, L.; Maheswaranathan, N.; Nayebi, A.; Ganguli, S.; Baccus, S.
\newblock Deep learning models of the retinal response to natural scenes.
\newblock  Advances in Neural Information Processing Systems,  2016, pp.
  1361--1369.

\bibitem[Mahuas \em{et~al.}(2020)Mahuas, Isacchini, Marre, Ferrari, and
  Mora]{Mahuas20}
Mahuas, G.; Isacchini, G.; Marre, O.; Ferrari, U.; Mora, T.
\newblock A new inference approach for training shallow and deep generalized
  linear models of noisy interacting neurons.
\newblock {\em Advances in neural information processing systems} {\bf 2020},
  {\em 33},~5070--5080.

\bibitem[Goldin \em{et~al.}(2021)Goldin, Lefebvre, Virgili, Ecker, Mora,
  Ferrari, and Marre]{Goldin21}
Goldin, M.A.; Lefebvre, B.; Virgili, S.; Ecker, A.; Mora, T.; Ferrari, U.;
  Marre, O.
\newblock Context-dependent selectivity to natural scenes in the retina.
\newblock {\em bioRxiv} {\bf 2021}.

\end{thebibliography}
\end{document}